# Quenching of fluorescence of aromatic molecules by graphene due to electron transfer

H S S Ramakrishna Matte, K S Subrahmanyam, K Venkata Rao, Subi J George and C. N. R. Rao*

New chemistry Unit, Chemistry and Physics of materials Unit  and International Centre for Materials

Science,   Jawaharlal Nehru Centre for Advanced Scientific Research, Jakkur P. O., Bangalore  560 064

(India).

**Abstract:** Investigations on the fluorescence quenching of graphene have been carried out with two organic donor molecules, pyrene butanaoic acid succinimidyl ester (**PyBS, I** ) and oligo(*p*-phenylenevinylene) methyl ester (**OPV-ester, II** ). Absorption and photoluminescence spectra of **I** and **II** recorded in mixture with increasing the concentrations of graphene showed no change in the former, but remarkable quenching of fluorescence. The property of graphene to quench fluorescence of these aromatic molecules is shown to be associated with photo-induced electron transfer, on the basis of fluorescence decay and time-resolved transient absorption spectroscopic measurements.

* For correspondence: cnrrao@jncasr.ac.in, Fax: (+91) 80-2208 2760

**Introduction**

Graphene is a fascinating material with unusual electronic, magnetic, optical and thermal properties.[1] Interaction of electron-donor and -acceptor molecules with graphenes has been exploited to modify the electronic properties of graphene through ground-state charge-transfer.[2] Fluorescence quenching property of graphene has been made use for the selective detection of biomolecules[3] and other purposes.[4,5] Excited-state photophysical processes between nanocarbons such as SWNTs and fullerenes with donor and acceptor molecules have been of much interest, because of their relevance to photovoltaics.[6,7] Thus, adducts of SWNTs with pyrene and porphyrin are shown to exhibit fast electron transfer leading to long-lived charge separated states.[8] Polycyclic aromatic molecules such as perylene and coronene are known to interact with graphene enabling non-covalent functionalization of graphene,[9] but photophysical aspects of such systems have not been explored. Quenching of the fluorescence of porphyrin by graphene and photophysical properties of porphyrin-graphene complexes have been reported,[10] but the mechanism of quenching of fluorescence by graphene has not been examined adequately, although it is believed that electron or energy transfer between graphene and the aromatic molecule would be responsible for the phenomenon. Theoretical studies show that long-range energy transfer is operative in the fluorescence quenching of a dye molecule in the presence of graphene.[11] The quenching of the green emission of ZnO nanoparticles accompanying the photoreduction of graphene oxide is, however caused by electron transfer from ZnO. Electron transfer has been similarly invoked in the case of $TiO_2$- graphene oxide. [4b] We were interested in investigating the mechanism of quenching the fluorescence of aromatic molecules by graphene via electron or energy transfer. With this purpose, we have investigated the quenching of fluorescence of two aromatic molecules by non-covalent interaction with graphene and demonstrated the occurrence of intermolecular photo-induced electron transfer. Specifically, we have examined the interaction of graphene with pyrene-butanaoic acid succinimidyl ester, (**PyBS**), **I** , and oligo(*p*-phenylenevinylene) methyl ester (**OPV-ester**), **II** ,  shown in scheme 1 with a graphene derivative, EGA, soluble in chloroform and dimethylformamide (DMF).

## Experimental section

The graphene derivative soluble in chloroform and DMF was prepared as follows. Graphite oxide (GO), synthesized by employing the literature procedure, [12] was subjected to thermal exfoliation in a furnace preheated to 1050 ºC under argon flow to obtain few-layer graphene (EG). The graphene sample contains 5 layers as determined by transmission electron microscopy (TEM) and atomic force microscopy. Conc. nitric acid (2 ml), conc. sulfuric acid (2 ml) and water (16 ml) were added to 50 mg of EG and the mixture heated in a microwave oven for 10 min under hydrothermal conditions [9a,12]. The product obtained after subsequent heating at 100 ºC for 12 h was washed with distilled water and centrifuged repeatedly to remove traces of acid. This product was refluxed with excess $SOCl_2$ for 12 h and the unreacted $SOCl_2$ removed under vacuum. The product was treated with dodecylamine (5 ml) under solvothermal conditions at 100 ºC. The amide-functionalized graphene (EGA) so obtained was characterized by IR spectroscopy and other means.[9a]

**PyBS**, **I**, was purchased from Sigma-Aldrich and the **OPV-ester, II** was synthesized according to the literature procedure.[13] UV-Vis spectra and fluoroscence spectra were recorded with a Perkin-Elmer Lambda 900 spectrometer and a Perkin-Elmer MPF 44B Fluorescence spectrophotometer respectively. Fluorescence decay was recorded in a time-correlated single-photon-counting spectrometer of Horiba-Jobin Yvon. Flash photolysis was carried out using a Nd:YAG laser source producing nanosecond pulses (8 ns) of 355 nm light with the energy of the laser pulse being around 200 mJ. Dichroic mirrors were used to separate the third harmonic from the second harmonic and the fundamental output of the Nd-YAG laser. The monitoring source was a 150 W pulsed xenon lamp, which was focused on the sample at 90° to the incident laser beam. The beam emerging through the sample was focused on to a Czerny-Turner monochromator using a pair of lenses. Detection was carried out using a Hamamatsu R-928 photomultiplier tube. Transient signals were captured with an Agilent infinium digital storage oscilloscope and the data was transferred to the computer for further analysis.

**Results and discussion**

Absorption spectra of **PyBS** , **I,** in DMF solution ($10^{-5}$ M) are shown in Figure 1 (a) in the presence of varying concentrations of graphene, EGA. The spectra show characteristic absorption bands of **I** around 314, 328 and 344 nm, intensities of which show an apparent increase with the EGA concentration. However, on subtracting the pyrene absorption, we obtain a broad band around 270 nm corresponding to the absorption band of graphene.[14] In the Figure 1 (b), we show the electronic absorption spectrum of **OPV ester**, **II** , in chlororform solution ($10^{-5}$ M) as a function of EGA concentration. The **OPV ester** exhibits characteristic absorption bands at 332 nm and 406 nm.[15] The increase in intensities of these bands with the graphene concentration is entirely accounted for the increasing intensity of the graphene absorption band around 270 nm. Thus, electronic absorption spectra of **I** + EGA and **II** + EGA show no evidence of interaction between the two molecules in the ground state. We also do not see of new absorption bands attributable to charge-transfer.

Unlike the absorption spectra, fluorescence spectra of **I** and **II** show remarkable changes on the addition of EGA. The intensity of the fluorescence bands decrease markedly with the increase in EGA concentration as illustrated in the Figure 2. We find similar fluorescence quenching with rhodamine B as well as coronene derivatives. Such strong quenching of fluorescence of aromatics by graphene can only be due to an excited state phenomenon since we do not observe ground state charge-transfer. Stern–Volmer plots based on the temperature variation study show a decrease in the quenching efficiency with increase in temperature suggests that mainly a static mechanism is operative,[16] possibly with a small dynamic contribution since the plots show a slight upward curvature.

Fluorescence decay measurements on **I** monitored at 395 nm could be fitted to a three-exponential decay[17] with lifetimes of 1.8, 5.7 and 38.7 ns. Addition of EGA causes a significant decrease in all the three lifetimes (Figure 3) with values 1.2, 4.6 and 29.1 ns respectively for the addition of 0.3 mg of EGA. A similar decrease in lifetime was observed in the case of **II** as well, though the actual decay profile showed slightly different features. Such multiexponential decay has been reported in the case of pyrene and porphyrin derivatives.[17, 18a]

Considering that the observed fluorescence quenching and life-time changes could due to an excited state photo-induced energy or electron transfer, we have carried out laser flash photolysis studies to explore the transient species. In Figure 4, we compare the transient absorption spectrum of the pure **I** with that of **I** on addition of 0.3 mg of graphene. The spectrum of **PyBS** shows an absorption maximum around 430 nm together with a broad band in the 450- 530 nm range due to the triplet state.[19] Upon addition of EGA, new bands emerge around 470 and 520 nm in the transient absorption spectrum at 500 nanoseconds. The 470 nm band can be assigned to the pyrenyl radical cation as reported in the literature,[17] suggesting the occurrence of photo-induced electron transfer from the **PyBS** to the graphene.

Accordingly, we observe the transient absorption around 520 nm which we assign to the graphene radical anion. The decay of the radical cation formed in the presence of graphene was fast, as evidenced from the appearance of a short-lived component (900 ns) in the decay profile (Figure 5). However, the decay of the transient absorption of pure **PyBS** monitored at 470 nm (see inset of Figure 5) shows a long-lived triplet with a lifetime of 6.17 microseconds. The transient absorption at 520 nm decays simultaneously with that of the pyrene radical cation indicating that it is due to the graphene radical anion. The bi-exponential nature of the decay after the addition of graphene was observed even at longer wavelengths suggesting a broad absorption for the graphene radical anion with maximum at 520 nm, as observed in the transient spectrum

In order to investigate the intermolecular electron transfer from **II** to EGA, pico- second laser flash photolysis was performed on a mixture of EGA and the OPV ester. In Figure 6, we have shown the transient absorption spectra of **II** with EGA at 1500 ps. Transient absorption spectrum of pure OPV showed maxima around 600 nm which could be attributed to the triplet state of the OPV following the literature[14a]. The transient absorption observed at 690 nm is considered to be due to a singlet-singlet transition. Most importantly, the transient spectrum at shorter time scales shows a band centered at $\lambda = 520$ nm which we attribute to the graphene radical anion just as in the case of **I**. The absorption band of the OPV radical cation appears in the Near-infrared region (1450 nm). [14] [b] Even though the main mechanism is electron-transfer in our systems, there might be chance of having energy transfer [11] as well, where both process are reported in some instances.

**Conclusions**

In summary, we observe efficient quenching of fluorescence of aromatic molecules by graphene. Our studies confirm the occurrence of photo-induced electron transfer between the aromatic molecules and graphene. Such electron-transfer has been observed in porphyrin-fullerene/nanohorn[18] and pyrene-carbon nanotube systems.[17] Occurrence of both electron transfer and energy transfer has been invoked by some workers to explain fluorescence quenching [20] and we are not able to entirely preclude the occurrence of energy transfer in the systems studied by us. The charge separated species reported by us in which graphene acts as an acceptor are long-lived. This feature could be of value in the design of photovoltaics.

**Acknowledgement:** The authors thank Prof. P. Ramamurthy, National centre for Ultra fast process (NCUFP) Chennai, for lifetime and Transient absorbtion spectroscopic studies.

.

Figure captions:

Figure 1: Electronic absorption spectra of (a) **PyBS, I,** ( $10^{-5}$ M in DMF) and (b) **OPV ester**, **II**, ($10^{-5}$ M in chloroform) with increasing concentration of graphene (EGA).

Figure 2: Fluorescence spectra of (a) **PyBS, I,** ( $10^{-5}$ M in DMF) and (b) **OPV ester**, **II**, ($10^{-5}$ M in chloroform) with increasing concentration of graphene (EGA).

Figure 3: Effect of addition of graphene (EGA) on the fluorescence decay of **PyBS, I,** ($10^{-5}$ M in DMF) at room temperature.

Figure 4: Effect of addition of EGA on the transient absorption spectrum of **PyBS**, **I**, ($\lambda_{exc}$ = 355 nm) after 500 ns.

Figure 5: Life time decay of transient species of **PyBS** + EGA recorded at 470 and 520 nm. Inset shows the decay of pure **PyBS** at 470 nm.

Figure 6: The transient absorption spectrum of **OPV ester** + EGA ($\lambda_{exc}$ = 355 nm) after 1500 ps.

**Table of Contents**

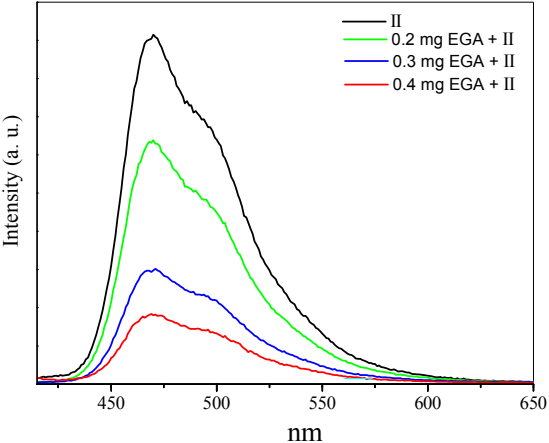

Figures

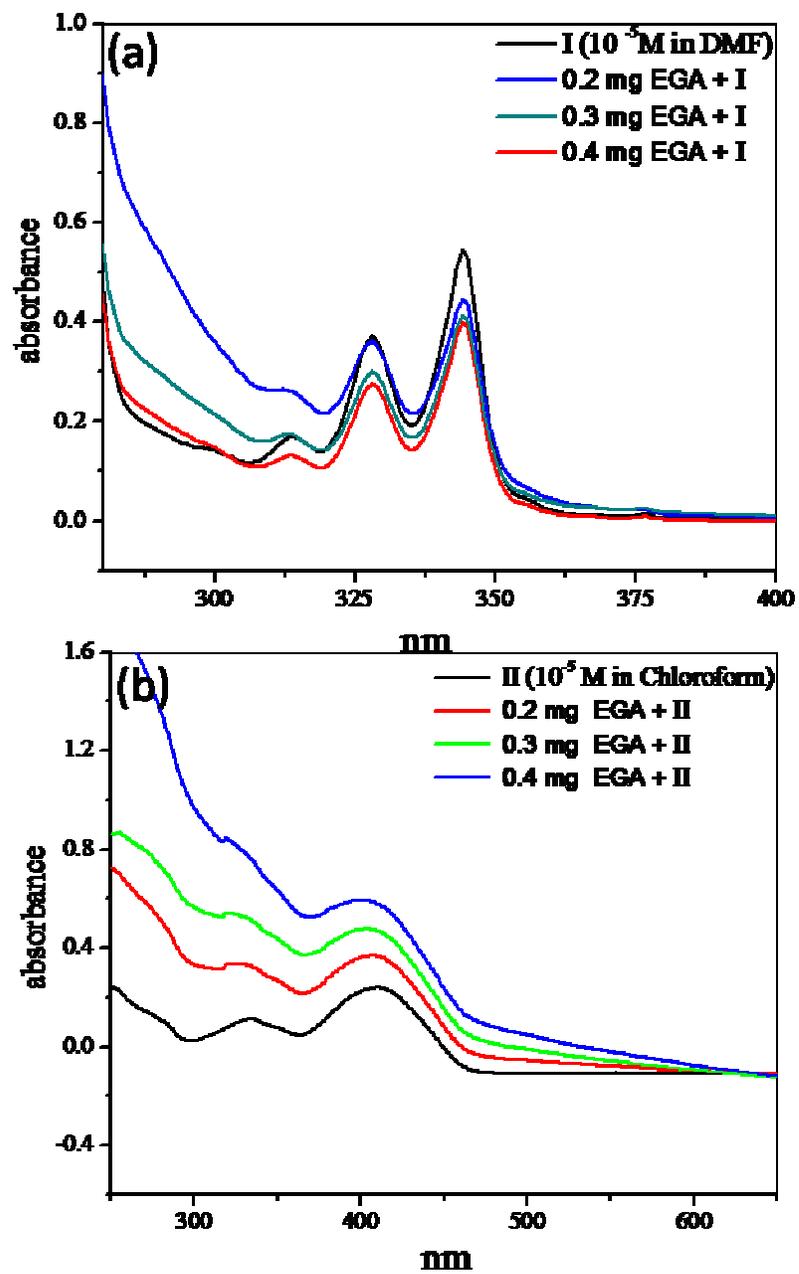

Figure 1

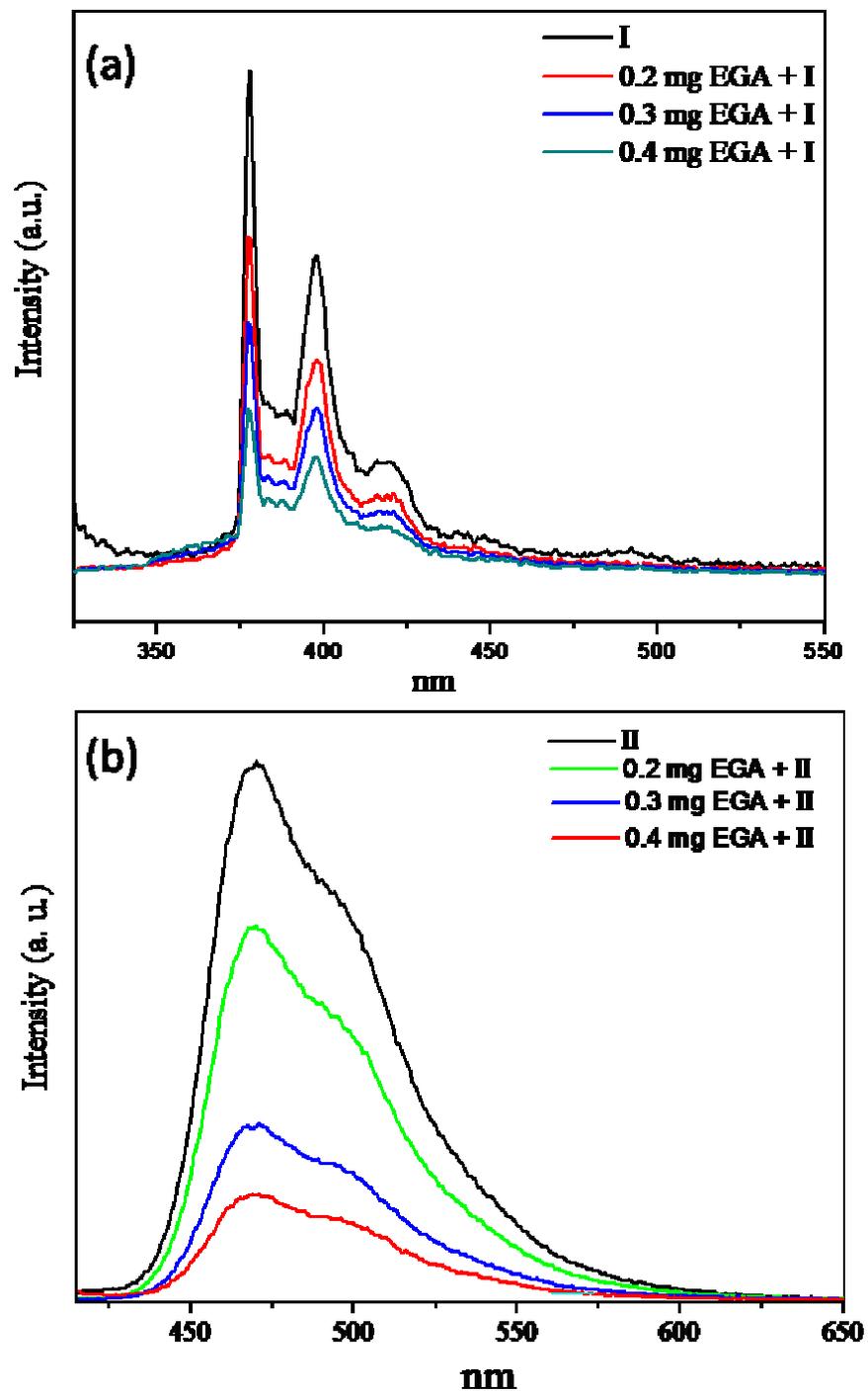

Figure 2

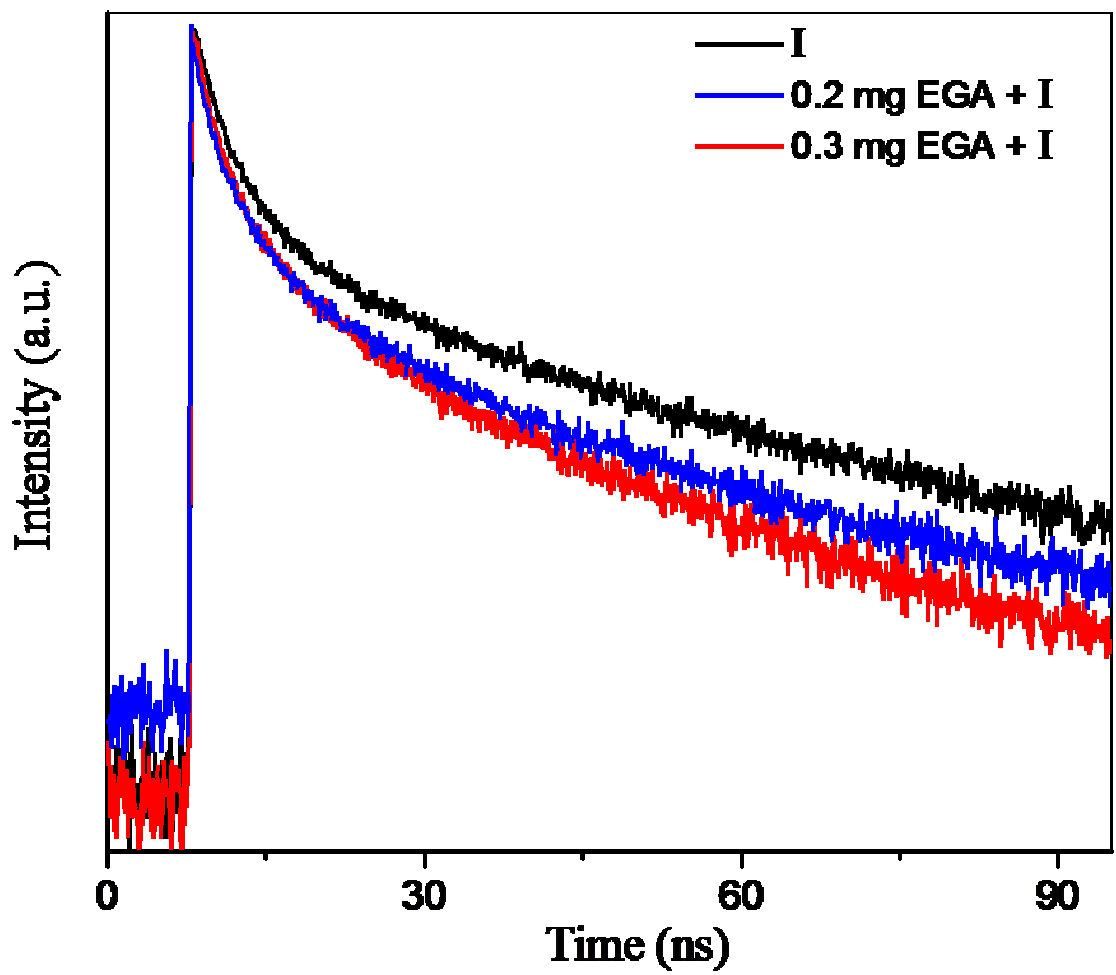

Figure 3

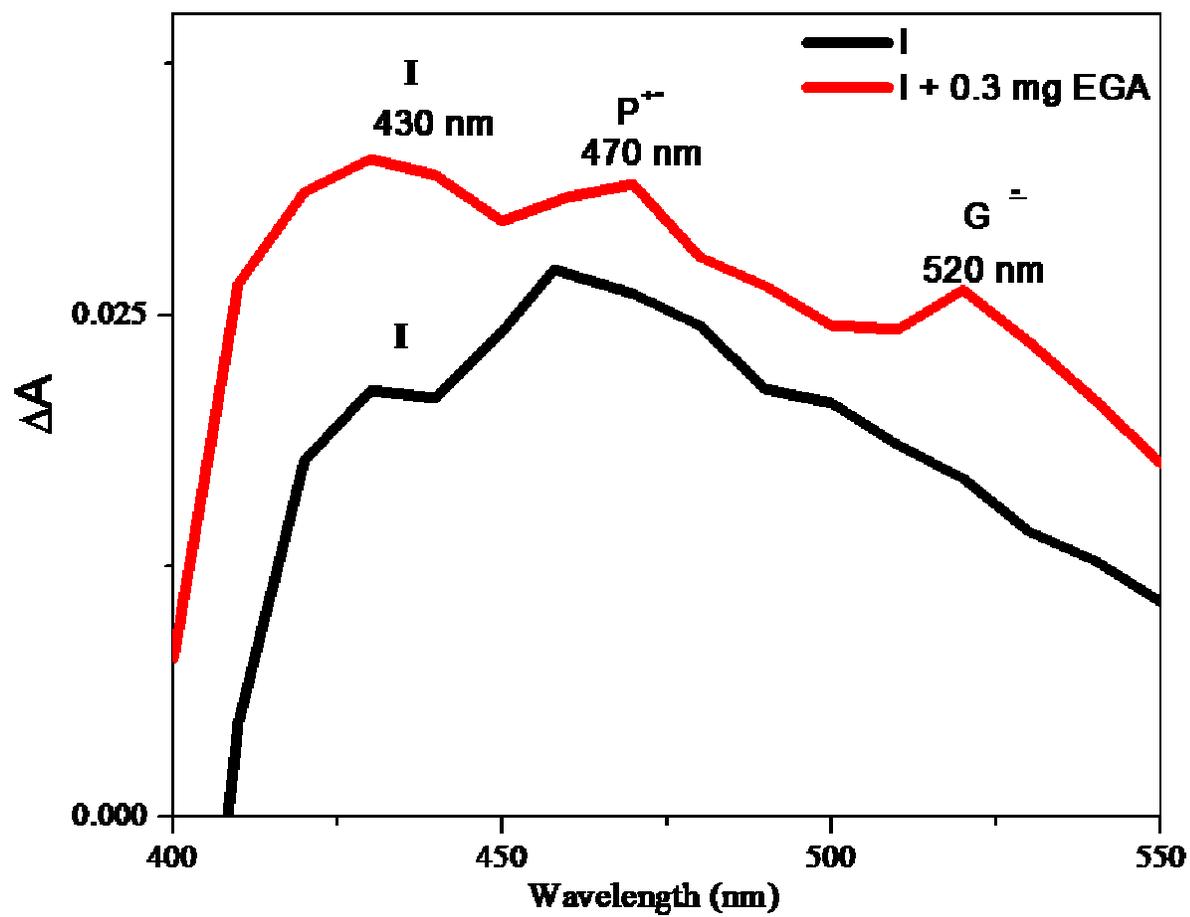

Figure 4

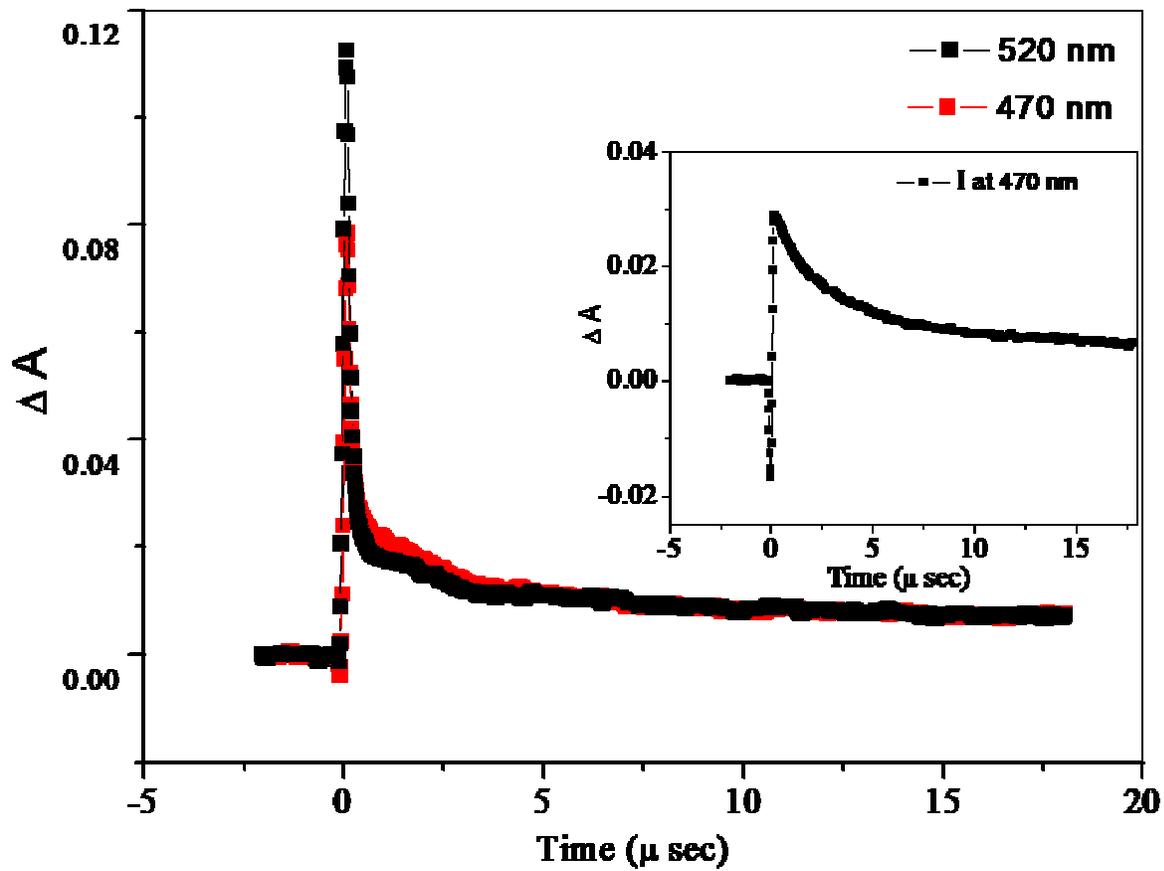

Figure 5

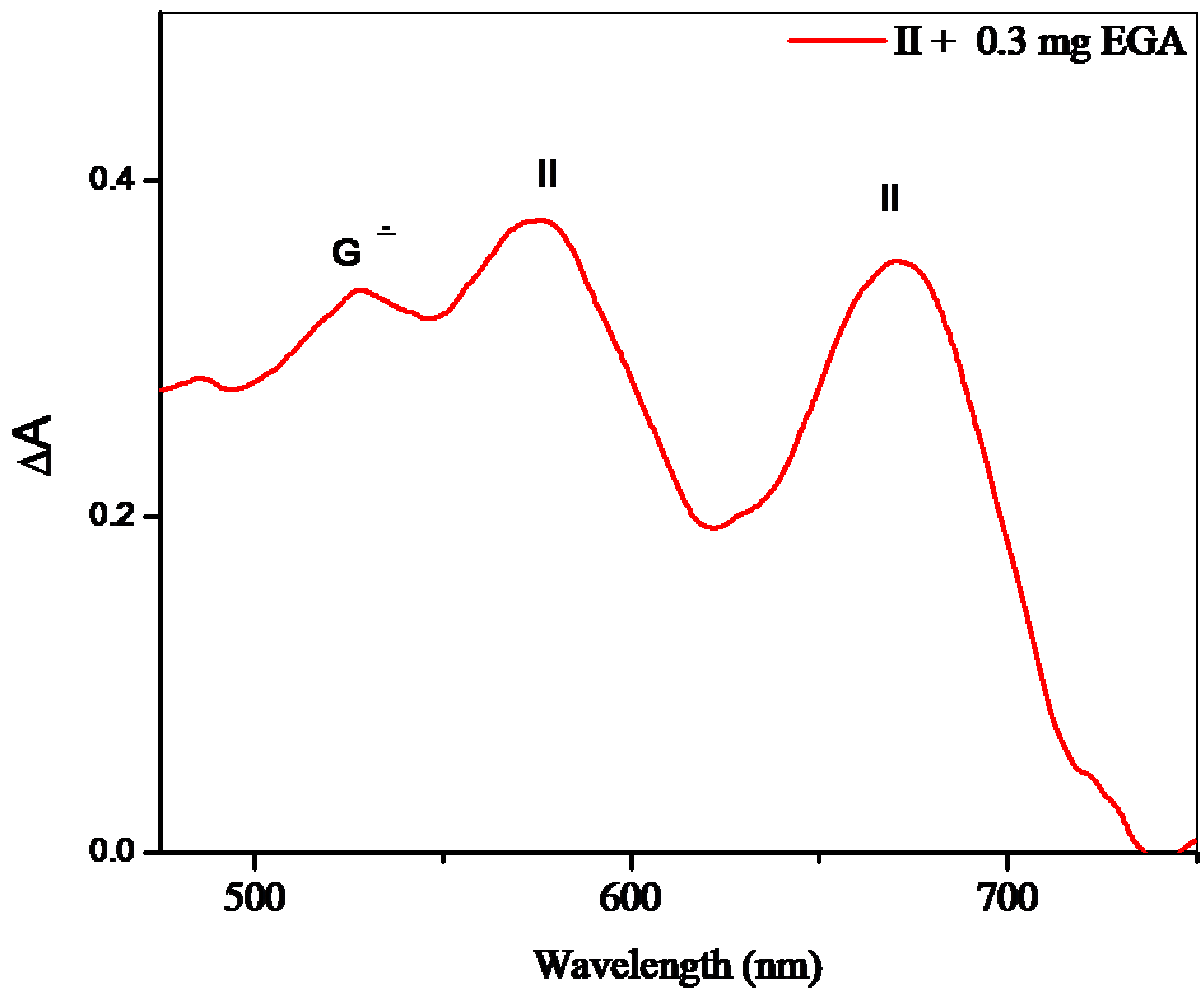

Figure 6

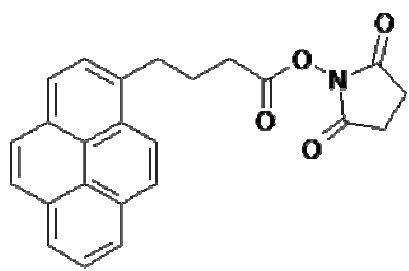

(**I**)

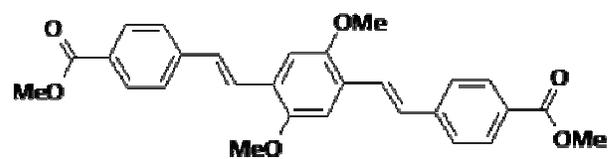

(**II**)

Scheme 1